\documentclass[10pt,twocolumn,letterpaper]{article}

\usepackage{cvpr}
\usepackage{times}
\usepackage{epsfig}
\usepackage{graphicx}
\usepackage{amsmath}
\usepackage{amssymb}
\usepackage{subcaption}

\usepackage[pagebackref=true,breaklinks=true,letterpaper=true,colorlinks,bookmarks=false]{hyperref}

\cvprfinalcopy 

\def\cvprPaperID{****} 

\ifcvprfinal\fi
\begin{document}

\title{GnetDet: Object Detection Optimized\\ on a 224mW CNN Accelerator Chip at the Speed of 106FPS}

\author{Baohua Sun, Tao Zhang, Jiapeng Su, Hao Sha\\
Gyrfalcon Technology Inc.\\
1900 McCarthy Blvd Suite 208, Milpitas, CA, 95035\\
{\tt\small baohua.sun@gyrfalcontech.com}
}

\maketitle

\begin{abstract}

Object detection is widely used on embedded devices. With the wide availability of CNN (Convolutional Neural Networks) accelerator chips, the object detection applications are expected to run with low power consumption, and high inference speed. In addition, the CPU load is expected to be as low as possible for a CNN accelerator chip working as a co-processor with a host CPU. In this paper, we optimize the object detection model on the CNN accelerator chip by minimizing the CPU load. The resulting model is called GnetDet. The experimental result shows that the GnetDet model running on a 224mW chip achieves the speed of 106FPS with excellent accuracy.

\end{abstract}


\begin{figure}
    \centering
    \begin{subfigure}[t]{0.23\textwidth}
        \centering
        \includegraphics[width=\linewidth]{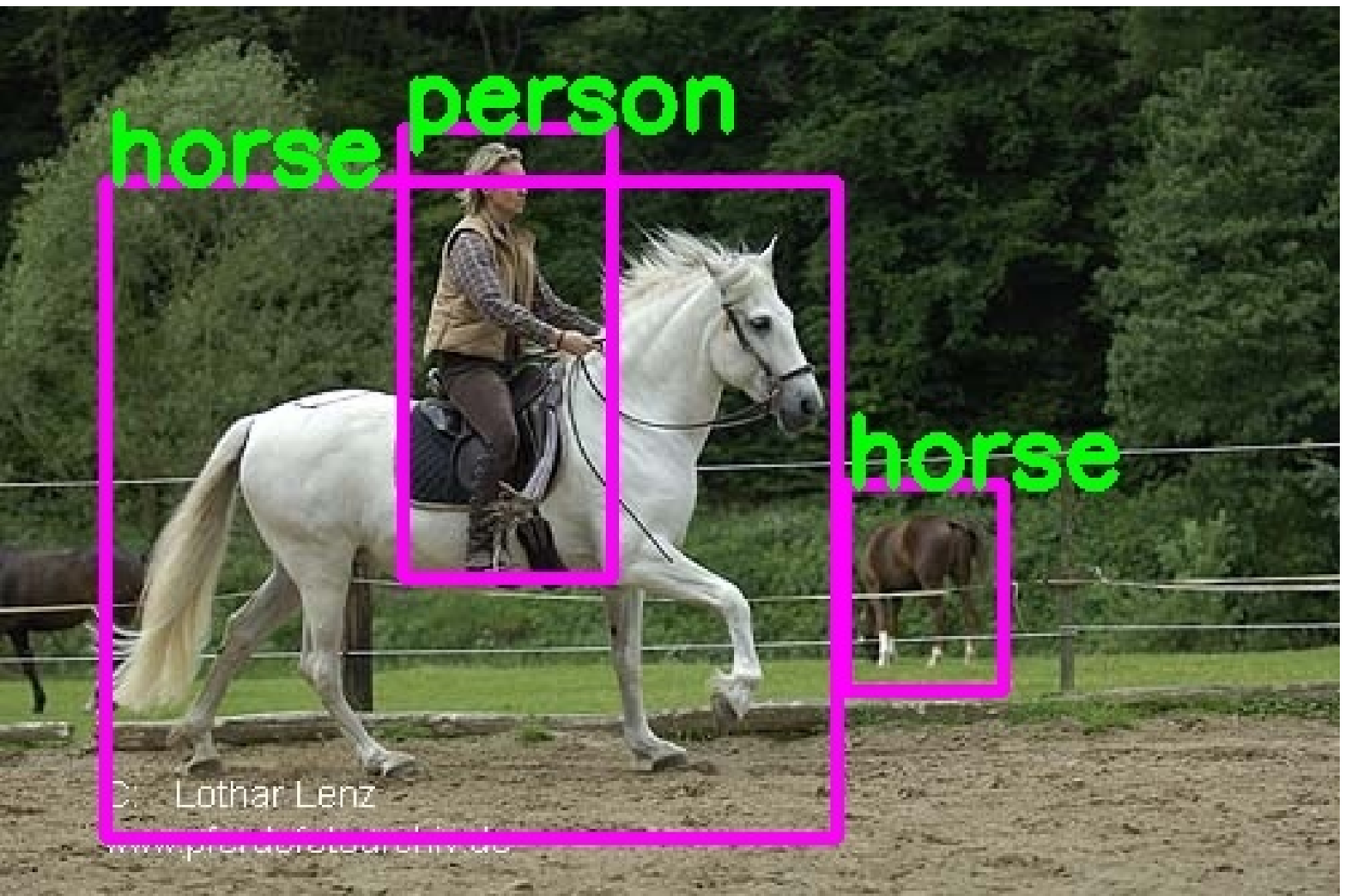} 
    \end{subfigure}
    \begin{subfigure}[t]{0.22\textwidth}
        \centering
        \includegraphics[width=\linewidth]{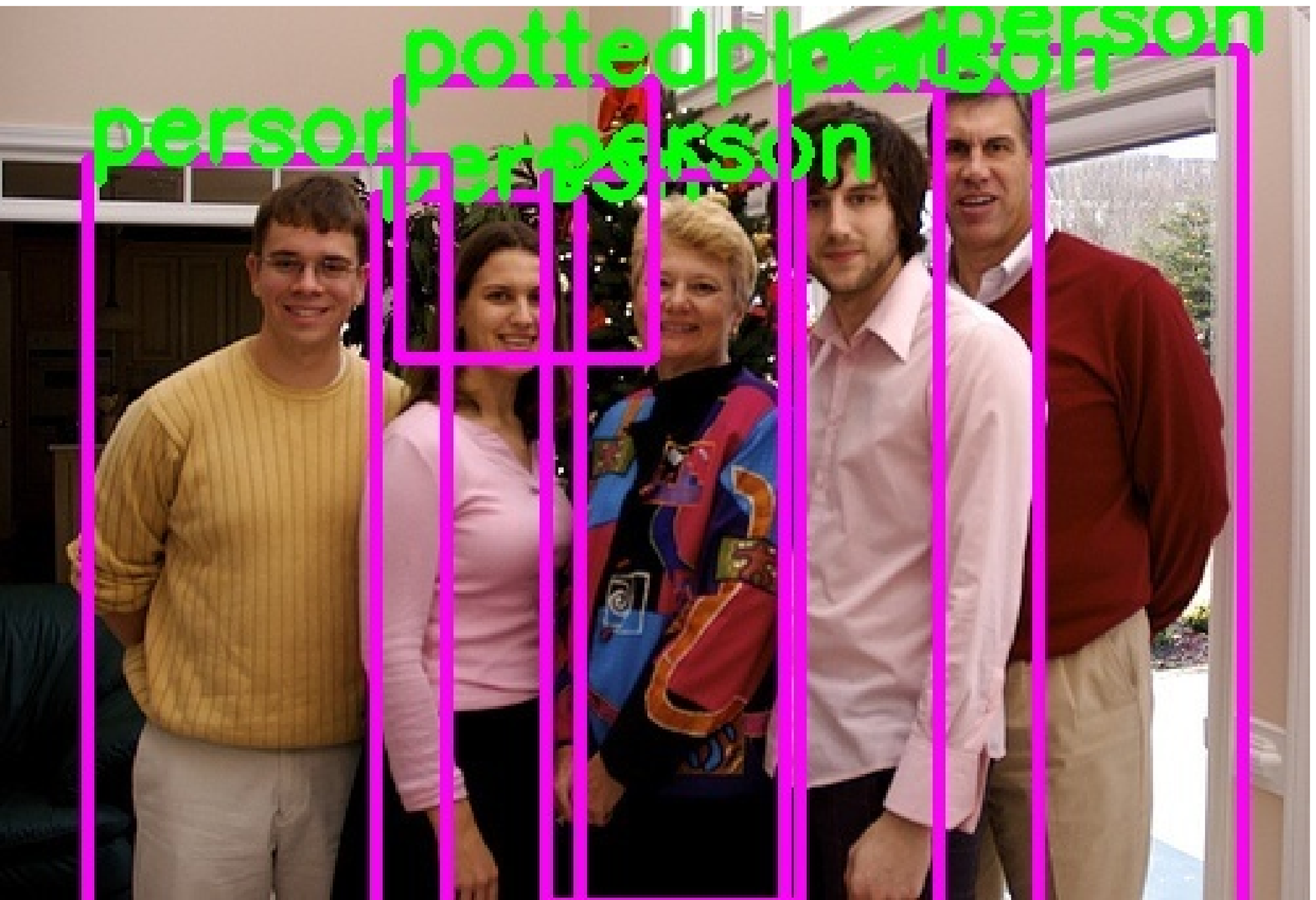} 
    \end{subfigure}
    \hfill
    \begin{subfigure}[t]{0.22\textwidth}
        \centering
        \includegraphics[width=\linewidth]{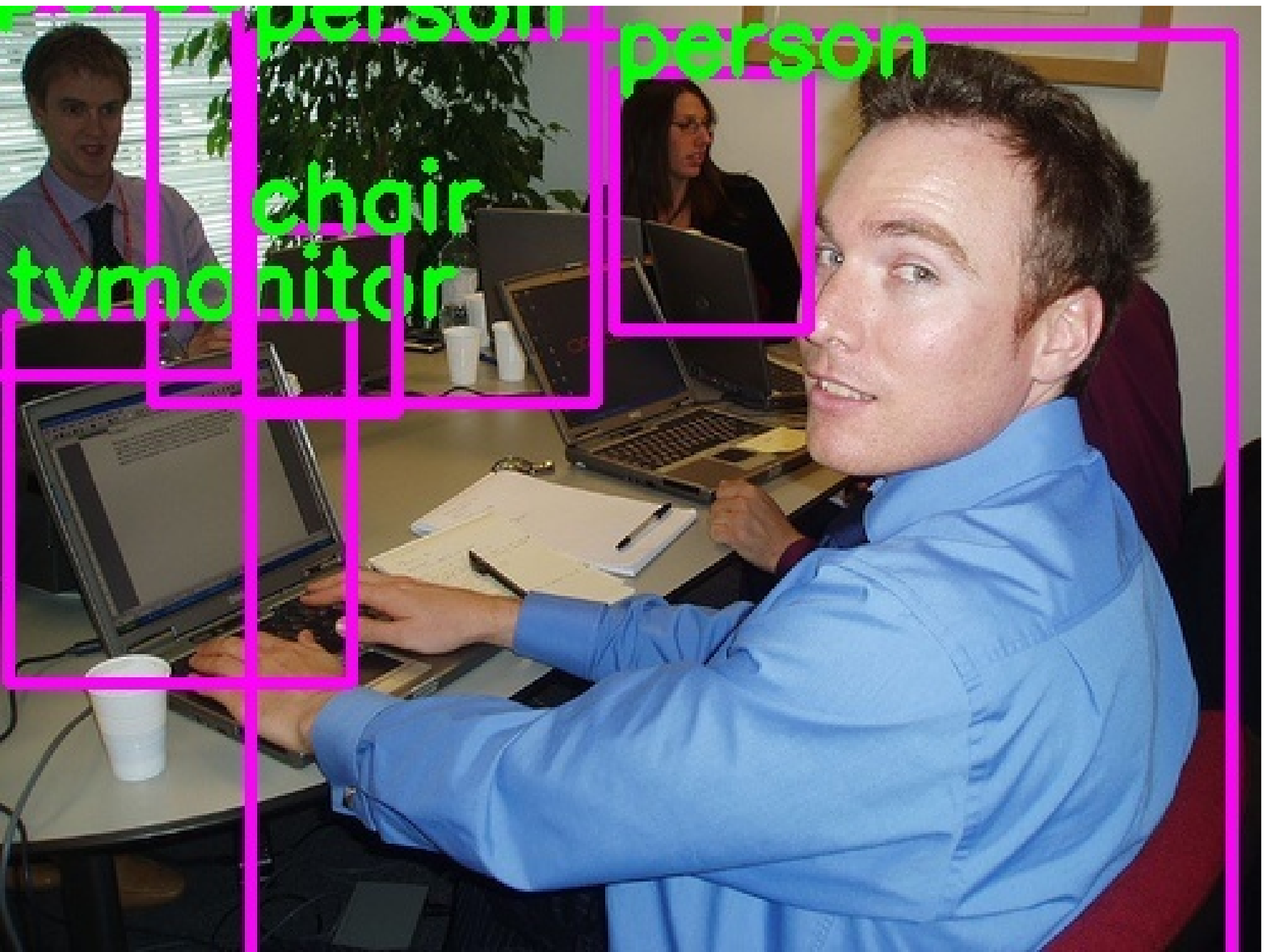} 
    \end{subfigure}
    \begin{subfigure}[t]{0.22\textwidth}
        \centering
        \includegraphics[width=\linewidth]{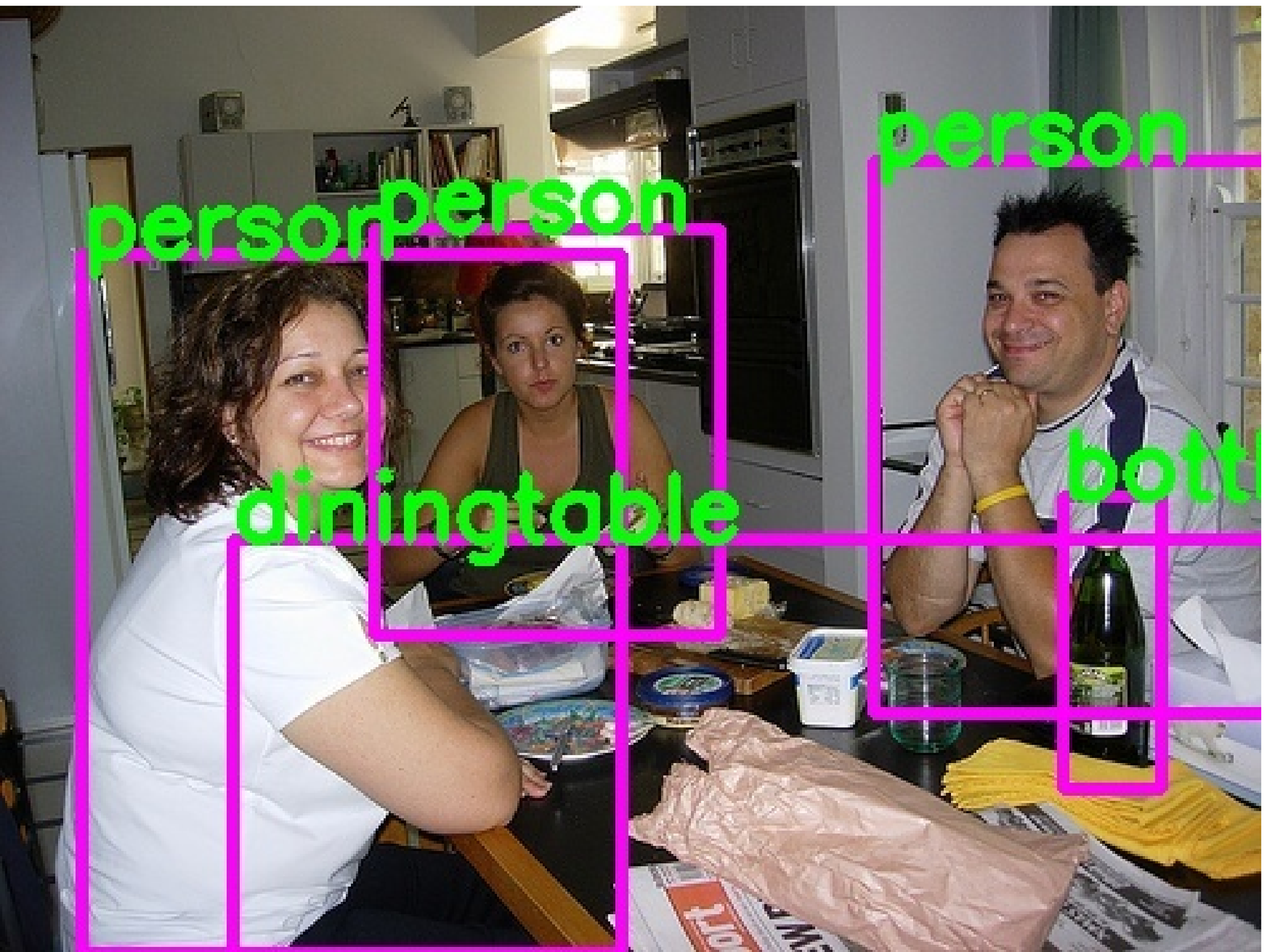} 
    \end{subfigure}
    \hfill
    \begin{subfigure}[t]{0.22\textwidth}
        \centering
        \includegraphics[width=\linewidth]{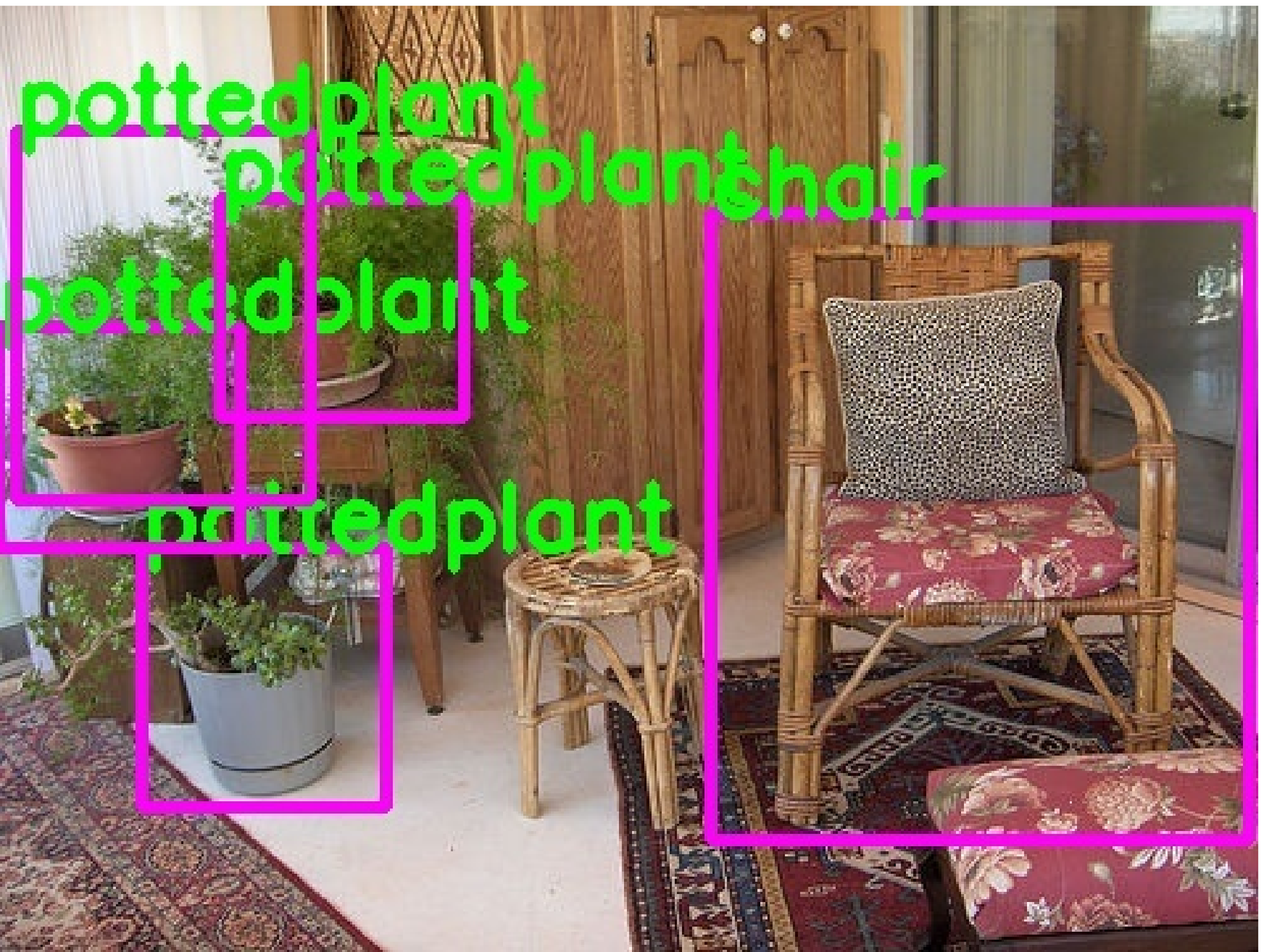} 
    \end{subfigure}
    \begin{subfigure}[t]{0.23\textwidth}
        \centering
        \includegraphics[width=\linewidth]{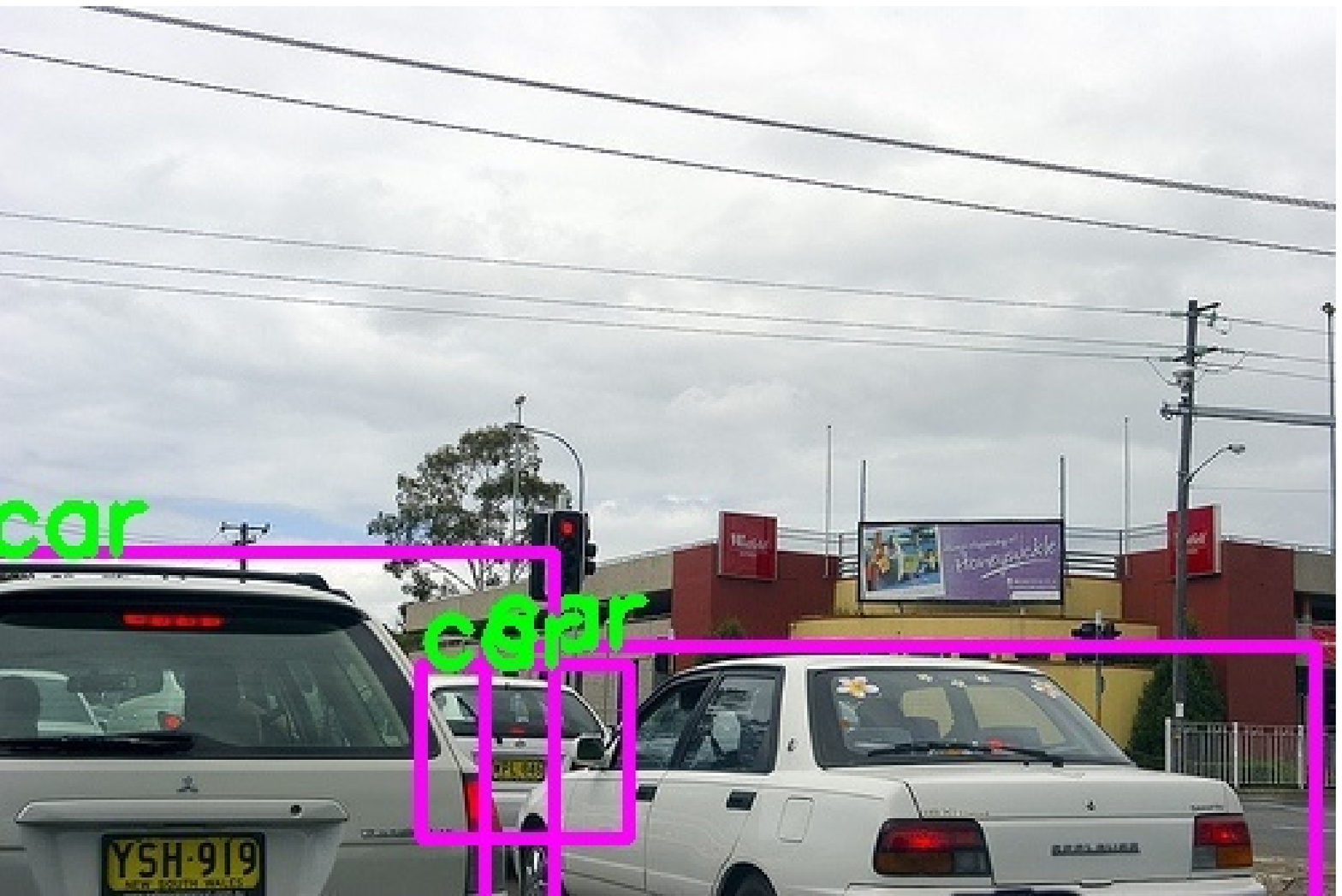} 
    \end{subfigure}
    \hfill
    \begin{subfigure}[t]{0.16\textwidth}
        \centering
        \includegraphics[width=\linewidth]{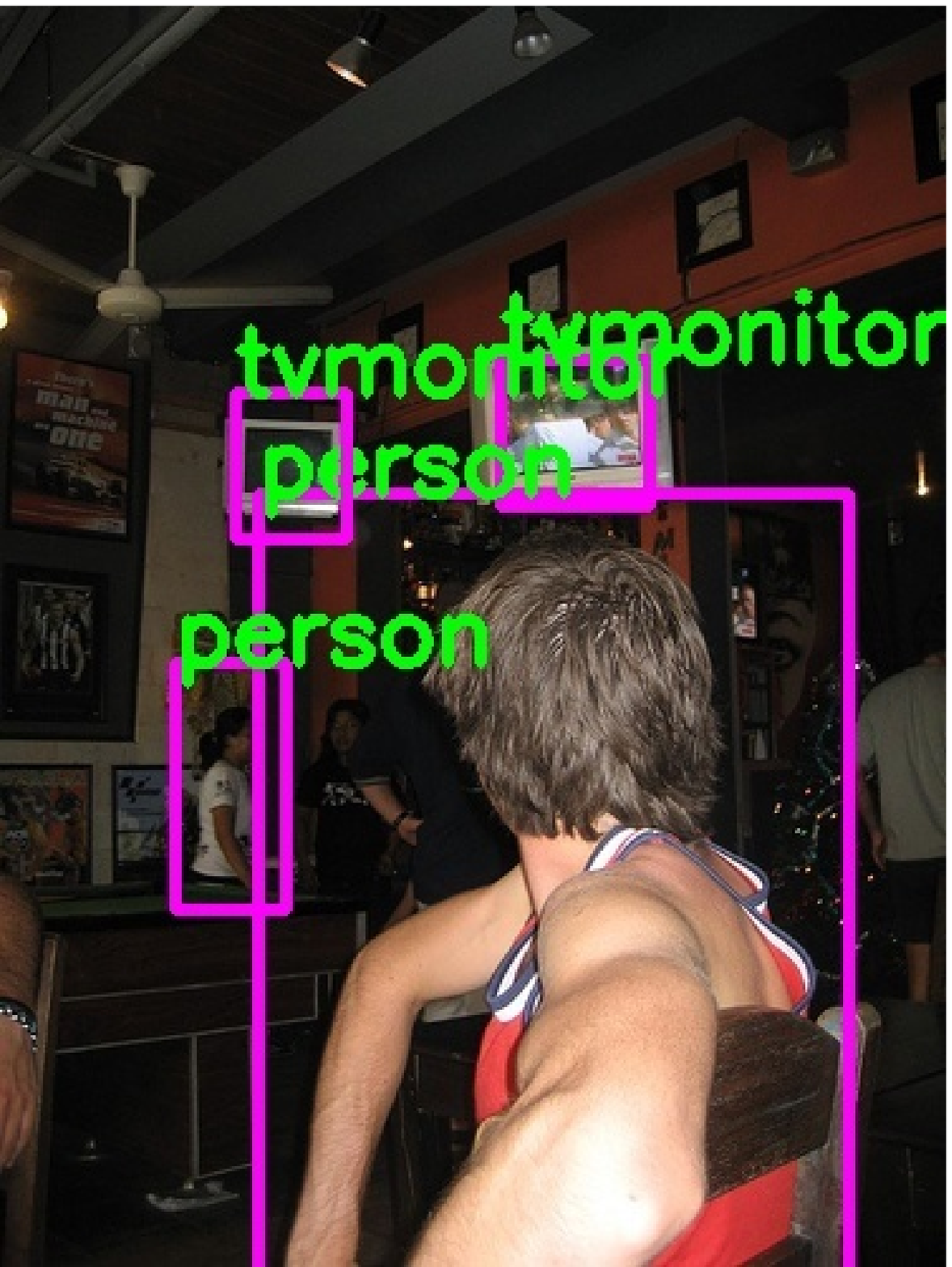} 
    \end{subfigure}
    \begin{subfigure}[t]{0.14\textwidth}
        \centering
        \includegraphics[width=\linewidth]{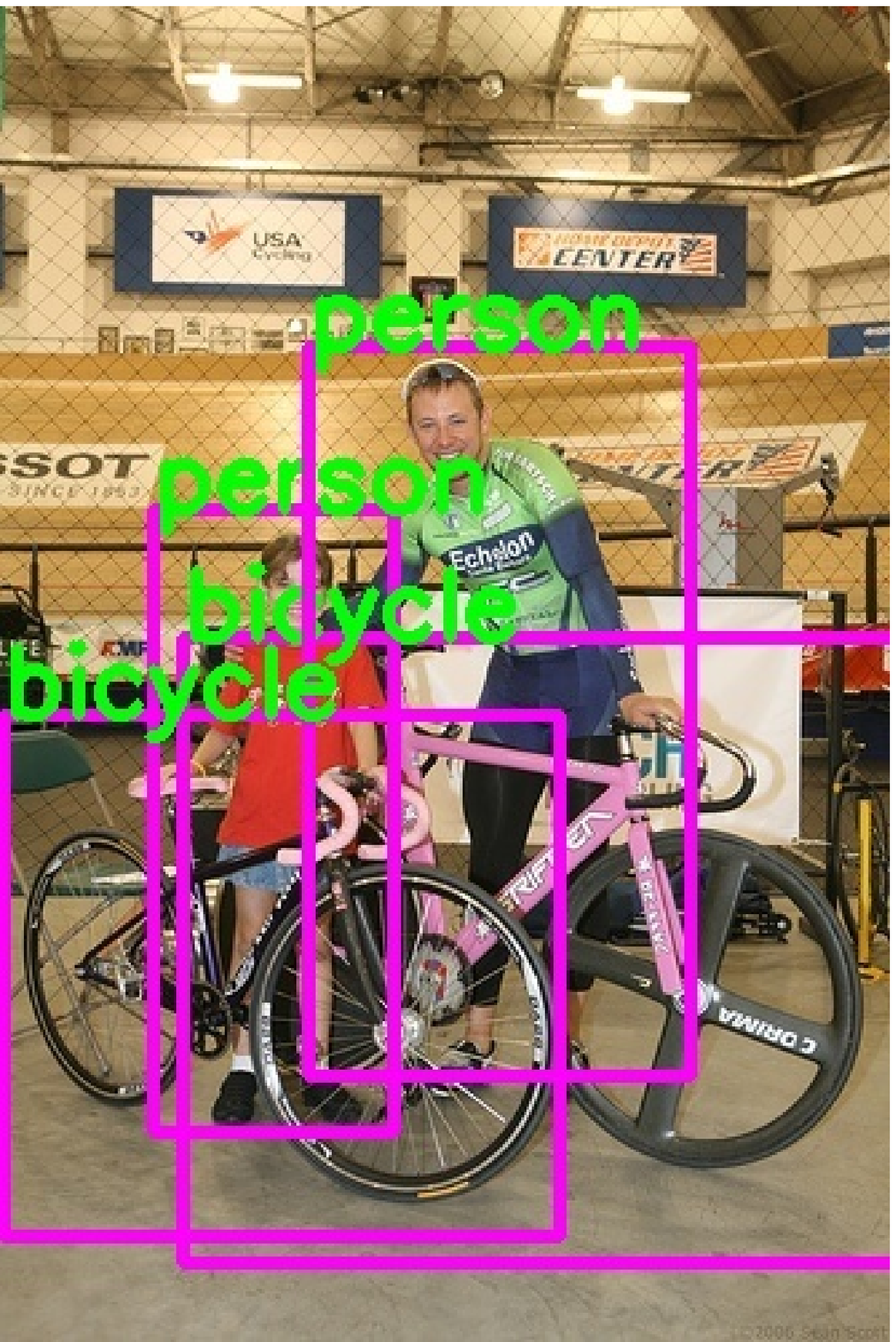} 
    \end{subfigure}
    \begin{subfigure}[t]{0.14\textwidth}
        \centering
        \includegraphics[width=\linewidth]{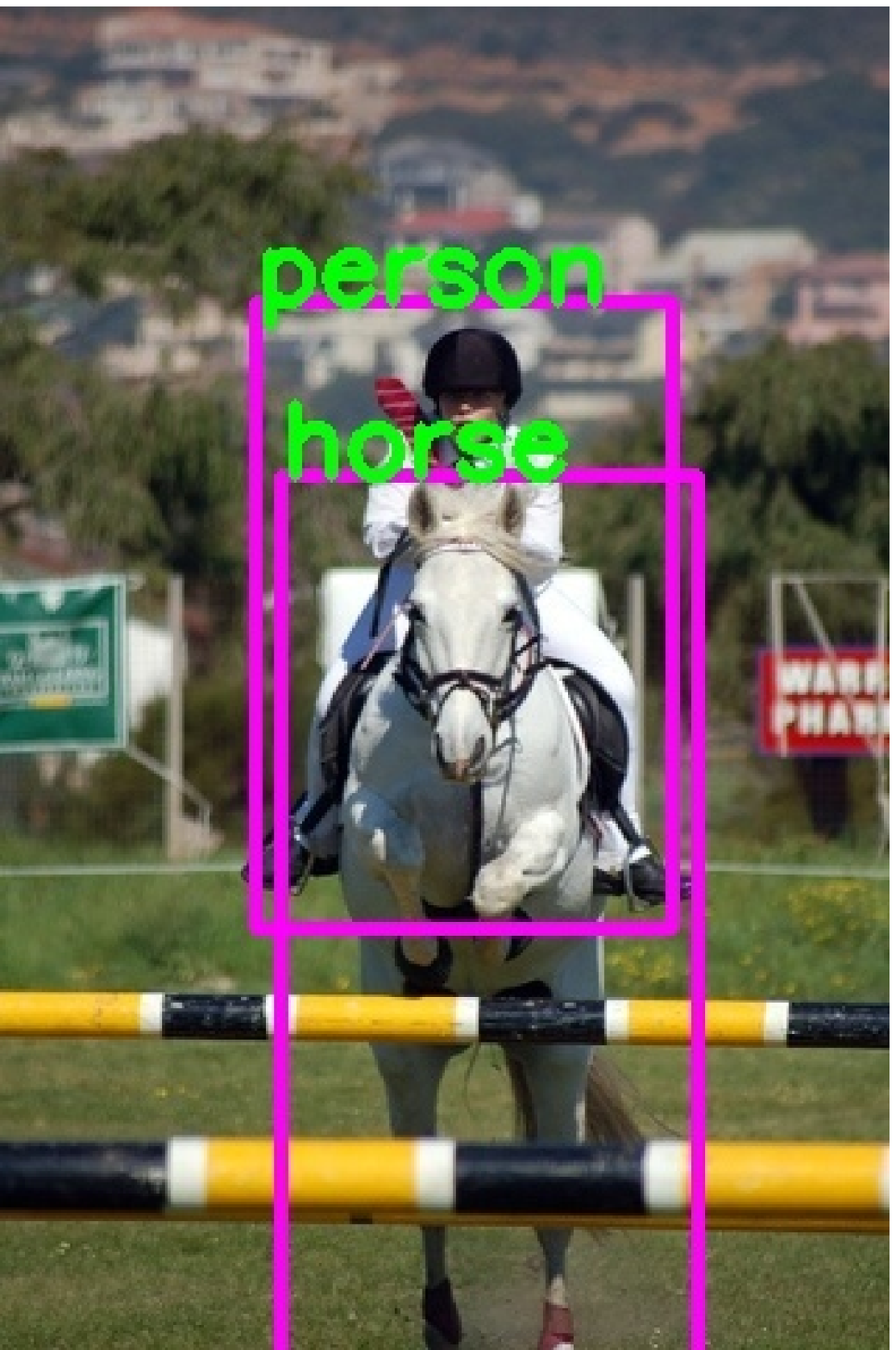} 
    \end{subfigure}
    \caption{Object detection results given by the proposed GnetDet model deployed on a 224mW accelerator chip.}
    \label{GnetDetExamples}
\end{figure}

\section{Introduction}

Object detection is a computer vision task to detect multiple objects in the input image. It has many real-world applications to be deployed on edge devices. CNN (Convolutional Neural Networks) accelerators~\cite{sun2018ultra,sun2018mram} are ideal for these applications by providing high inference speed and low power consumption. The most recent chip has the peak power of only 224mW~\cite{sun2021gnetseg}. These low-power CNN accelerators are used in computer vision tasks~\cite{sun2020cdva,sun2019demonstration,sun2020superocr}, and also in NLP (Natural Language Processing) tasks~\cite{sun2018super,sun2019squared,sun2019superchat,sun2019supercaptioning,sun2019system,sha2019device}, and extended into tabular data machine learning~\cite{sun2019supertml} and multimodal tasks~\cite{sun2020multi}.

For real-world applications, the performance of power consumption, CPU load, inference speed, and prediction accuracy should be considered jointly. Minimizing the CPU load releases the CPU for other tasks, and also helps improve the inference speed. This paper proposes the GnetDet model for the object detection task, which is tailored to the CNN accelerator chip. It fully utilizes the computational resources on the CNN accelerator chip and only the NMS (Non Maximum Suppression) is running on the CPU. On a Raspberry Pi 3B with only USB2.0 interface, the proposed GnetDet model running on a 224mW CNN accelerator chip processes a single Y-channel 224x224 image at the speed of 66.67FPS(Frame Per Second). On a i5-7300HQ@2.5GHz CPU as the host CPU with USB3.0 interface, the speed is over 106FPS. Fig.~\ref{GnetDetExamples} shows some results given by the GnetDet model.

\section{Motivations from the Existing Work}
\subsection{From GnetFC-v1 to GnetFC-v2}
GnetFC-v1~\cite{sun2020superocr, sun2020cdva} is an image classification model specially designed for the CNN accelerator chip. It can be also used for NLP and multi-modal tasks~\cite{sun2019system,sun2019demonstration,sun2020multi}. The motivation of GnetFC-v1 model is to reduce the CPU load of image classification models by replacing the FC (Fully Connected) layers with the 3x3 convolutional layers. This is done by using multiple convolutional layers without padding at the end of the CNN model. For an input image with a resolution of 224x224, the feature map size of the VGG~\cite{simonyan2014very} convolutional layers output is 7x7. After three layers of convolutions without padding, the output size will become 1x1. And the number of channels will be used to represent the model scores for all the classes, which can be used to obtain the classification result.

The GnetFC-v1 successfully reduces the CPU load to zero for the FC layers. However, the maximum number of classes is capped by the maximum number of channels at the last layer of the GnetFC-v1 model due to the limited memory on the CNN accelerator chip. For image classification with large number of classes, such as ImageNet~\cite{deng2009imagenet} with 1000 classes, the GnetFC-v1 model is infeasible to be implemented on the CNN accelerator chip. 

We propose the GnetFC-v2 model to address this problem. The main problem of GnetFC-v1 is inefficiently utilizing the entire feature map of each channel on the CNN accelerator chip. In GnetFC-v2, each pixel on the feature map of 7x7 is used as a classification score connected to a softmax layer. This immediately increase the number of classes for the image classification model by 49x. For example, a tensor with 256-channel feature maps can represent up to 12,544 classes, and a 512 channel will have that number doubled, which is more than 25x for the 1000 classes in ImageNet. Furthermore, if a 14x14 feature map is used to represent the classification scores, the number of classes can be further increased by 4x. That means for a 512 channel, the GnetFC-v2 model can support up to 147,968 classes, which is sufficient for the majority of image classification problems with large number of classes. 

The proposed GnetFC-v2 model is trained using ImageNet2012 dataset~\cite{ILSVRC15} with 1000 classes. After 5 epochs of training, the accuracy on the training dataset is 69.3\%, and the accuracy on validation dataset (50,000 images) is 60.0\% on the CNN accelerator chip. More training can further improve the accuracy, which will be the future work. When connected to a host CPU through a USB3.0 interface, the speed of the GnetFC-v2 model on the CNN accelerator chip is 11.95ms (83.68FPS), which includes CNN processings on the CNN accelerator chip and I/O (Input/Output). The I/O includes sending the image to the CNN chip and reading the output tensor from the CNN chip.

\subsection{From GnetFC-v2 to GnetDet}
The GnetFC-v2 fully uses the output feature map for the image classification task. When using the CNN chip for the object detection task, the same idea can be used for designing the output tensors from the CNN chip to represent the desired format, such as bounding boxes, confidence scores, and classification scores.

Object detection models are usually composed of backbone, detection neck and detection head~\cite{bochkovskiy2020yolov4}. The CNN accelerator chip has low-power and fast-speed to extract the features of the input image, but it does not directly support architectures with skip connections, such as FPN~\cite{lin2017feature}. Thus it is straightforward to run the backbone on the CNN accelerator, and leave the remaining of the model to be processed by the host CPU. For example, running the SSD~\cite{liu2016ssd} backbone on the CNN chip, and running the detection head on the host CPU. However, the CPU load will be high, thus the FPS will be low. This is not a feasible solution on embeded systems with low-end CPUs.

Similar to GnetFC-v2, the GnetDet model optimizes the object detection model on the CNN accelerator chip, and the object detection result is represented by the output tensor from the chip. Thus no more computation is needed for the detection head on the host CPU. The only load for the host CPU is the NMS, the computation of which is not heavy compared to the CNN model.

\begin{figure*}[htb]
  \centering
  \centerline{\includegraphics[width=17.5cm]{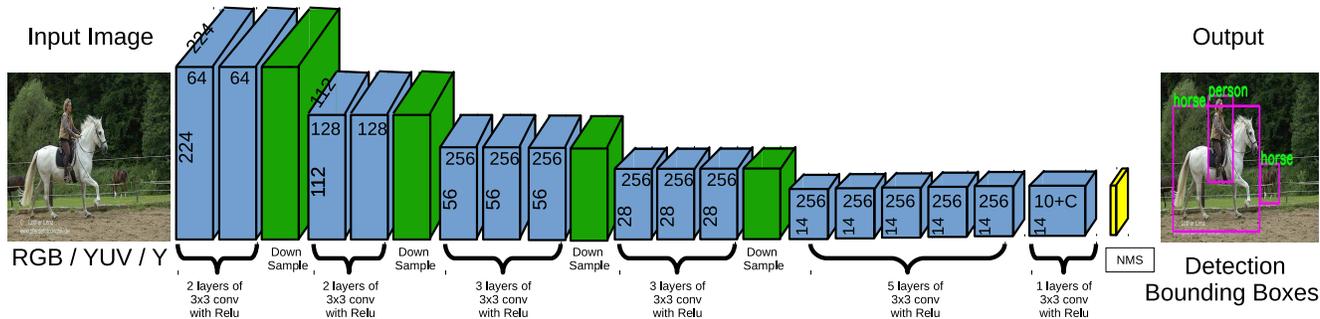}}
\caption{Illustration of the proposed GnetDet model architecture. Input image size of 224x224 is set as an example, while 448x448 is also supported. The input can be either 3-channels of RGB/YUV, or a single Y channel. The convolutional layers are composed by 3x3 convolution followed by Relu activation, which are separated by the down-samplings. The detection head has (10+$C$)x14x14 followed by the NMS block, where $C$ stands for the number of Classes. The entire model is deployed on the CNN accelerator except the NMS block. }
\label{GnetDetArchitecture}
\end{figure*}

\section{Proposed Model: GnetDet}
The proposed GnetDet model fully utilizes the memory and operators on the low-power CNN accelerator chip, and optimizes the joint performance metrics of CPU load, inference speed, and accuracy. The design of the proposed GnetDet model considers the factors including input format, model architecture, and output tensor representation. Each of these factors will be designed in detail in this section.

\subsection{Design of the Input Image Format}
Similar to the GnetSeg~\cite{sun2021gnetseg} model which is designed for semantic segmentation tasks on the CNN accelerator chip, the GnetDet model supports input image resolution of both 224x224 and 448x448, and supports RGB/YUV/Y format for the input channels.

\subsection{Design of the Model Architecture}
The proposed GnetDet model is composed of a backbone and a detection head that is similar to~\cite{redmon2016you}. However, there are three major differences. First, no FC layers inside the model. The CNN accelerator chip only supports convolutional operations, and the detection head is directly connected to the backbone through 3x3 convolutions without any FC layers. Second, only 3x3 conv and 2x2 max pooling are used in the GnetDet model. Third, the number of cells is 14x14 instead of 7x7, in order to improve the detection accuracy.

Depending on the detailed features of different chips, the model architecture of the convolutional layers may vary. We designed two different variants of the GnetDet model, namely GnetDet-Large and GnetDet-Small.

In the context of the CNN accelerator chips, the term of sublayer and major layer will be used. A sublayer is a convolutional layer followed by a Relu function. A major layer has multiple sublayers. Different major layers are separated either by a change of number of channels between sublayers, or by a change of the feature map size caused by downsampling. 

\subsubsection{GnetDet-Large}
GnetDet-Large has six major layers. The first five major layers serve as the backbone, which have a similar architecture to the convolutional layers of VGG except that the numbers of channels of the fourth and fifth major layers are reduced because of the memory constraint on the chip. The sixth major layer serves as the detection head to output the tensor that can be parsed as the bounding boxes, confidence score, and classification scores. The resulting model of GnetDet-Large is shown in Figure~\ref{GnetDetArchitecture}.

\subsubsection{GnetDet-Small}
GnetDet-Small has five major layers. Different from the above GnetDet-Large, the GnetDet-Small only has the first four major layers as the backbone. The fifth major layer merges the sixth major layer, so it has a total of six sublayers. The output tensor from the fifth major layer directly connects to the NMS.

\subsection{Design of the Output Tensor Representation}
The image is split into 14x14 cells. For each cell, there are two bounding boxes which uses 2x4 channels for the coordinates. The confidence scores for the two bounding boxes take another 2 channels. Thus, the output tensor is composed of 10+$C$ channels, where $C$ is the number of classes for the object detection task, which is similar to~\cite{redmon2016you}.

\section{Experiments}

\subsection{Datasets}
We used a combination of VOC2007~\cite{pascal-voc-2007}, VOC2012~\cite{pascal-voc-2012}, and MS-COCO2014~\cite{lin2014microsoft} datasets in our experiments. The 4,952 images in the VOC2007 validation dataset is used as our validation dataset in the experiment. The 5,011 images in the VOC2007 training dataset, and the 17,125 images in the VOC2012 training and validation dataset are used for training. In addition, the training and validation data in the COCO2014 dataset is also added to our training dataset. Since COCO2014 dataset has 80 classes, so only the data that overlaps with the VOC2017 20 classes are added in the training. A total of 53,891 images in COCO2014 training and validation dataset are added to our training dataset. Thus we have in total 76,027 images in our training dataset.

\begin{table*}
\begin{center}
\begin{tabular}{|c|c|c|c|c|}
\hline
Model  & host CPU & Interface & FPS & mAP \\
\hline
GnetDet-Large\_448\_RGB & i5-7300HQ@2.50GHz & USB3.0 &   63.70FPS & {\bf 0.6603}\\
GnetDet-Large\_224\_YUV & i5-7300HQ@2.50GHz & USB3.0 &  86.19FPS & 0.6420\\
\hline
GnetDet-Large\_448\_Y & i5-7300HQ@2.50GHz & USB3.0 &   85.62FPS & 0.6326\\
\hline
GnetDet-Large\_224\_Y & i5-7300HQ@2.50GHz & USB3.0 &   {\bf 106.18FPS} & 0.6202\\
GnetDet-Large\_224\_Y & RK3399@1.8GHz & USB3.0  &  100.15FPS & 0.6202\\
\hline
GnetDet-Large\_224\_Y & RK3399@1.8GHz & USB2.0 &  83.33FPS & 0.6202\\
GnetDet-Large\_224\_Y & Raspberry Pi 3B@1.2GHz & USB2.0 &   66.67FPS & 0.6202\\
\hline
\end{tabular}
\end{center}
\caption{Performance of GnetDet on the VOC2007 validation data. Only the NMS is running on the host CPU, and the CNN model is running on the CNN accelerator chip. The FPS is calculated by counting both the CNN processing time and the I/O time between the CNN accelerator chip and host CPU.}
\label{GnetDetmAPtable}
\end{table*}

\subsection{Experimental Results}
Fig.~\ref{GnetDetExamples} shows some examples of the object detection results given by GnetDet model implemented on the CNN accelerator chip.
 
Table~\ref{GnetDetmAPtable} shows the performance of GnetDet model deployed on the CNN accelerator chip connected to various host CPU platforms. We used three different platforms. The first one is a desktop which has i5-730000HQ CPU@2.5GHz with USB3.0 interfaces. The second one is an RK3399 having a Dual Core A72 and a Quad Core A53, with both USB2.0 and USB3.0 interfaces. The third one is a Raspberry Pi 3B (RBP3) which has a Quad Core A53 CPU and USB2.0 interface only. 

For the inference speed, the time for both I/O and CNN processing are counted. The GnetDet-Large\_224\_Y model running on the accelerator chip connected to an i5-7300HQ@2.50GHz with USB3.0 interface has the inference speed of more than 106FPS. On a RBP3 with USB2.0 interface, this model runs at the speed of 66.67FPS.

The comparison of the FPS metric between USB2.0 and USB3.0 interface on RK3399 shows that the I/O is the bottleneck for inference speed. For USB3.0 on RK3399, the speed for inference and I/O is 100.15FPS. While this number drops to 83.33 when using USB2.0. The comparison of the FPS metric between GnetDet-Large\_224\_YUV and GnetDet-Large\_224\_Y on i5-7300HQ@2.50GHz also proved this. The computation complexity of these two models are very similar, but the data transferred from the host CPU to the CNN chip on the former model is three times as much as the later model. This further proves that the bottleneck for speed is the I/O interface, but not the CPU. If the host processor has an even faster interface, such as PCIe, the speed could be even faster.

Despite the CNN model of the detector can be fully deployed on the chip, the post-processing of NMS and the pre-processings such as resizing the image and sending the resized image to the chip still runs on the host CPU. The parallel pre/post-processings on the host CPU and the GnetDet model on the CNN accelerator chip can further improve the speed of the whole system, which is out of the scope of this paper.

\section{Further Improvement and Future Work}
\subsection{GnetDet Accuracy Improvement}
In order to further improve the accuracy, more data can be added for training. We observe significant accuracy improvement on the VOC2007 validation dataset if we add more training data. If only VOC2007-train and VOC2012-trainval dataset are used for training, the detection accuracy on VOC2007-val is 61.18\% for 224 YUV input. When we add the COCO2014 data, the accuracy is improved by more than 3\% to 64.20\%. 

The accuracy of GnetDet model can be further improved by increasing the number of predefined anchors and clustering the predefined these anchors as done in YOLOv3~\cite{redmon2018yolov3}, and using data augmentations and various loss functions as listed in YOLOv4~\cite{bochkovskiy2020yolov4}, which will not be expanded in this paper.

\subsection{GnetDet without NMS}
For the GnetDet model, all the convolutional computations are running on the low-power and high-speed CNN accelerator. The only CPU load is for the NMS block where the computation is not intensive. However, for scenarios where CPU load is very sensitive, GnetDet can be further modified to remove this NMS block. Recent work~\cite{sun2020onenet} has successfully built end-to-end object detection model without the NMS block, with negligible accuracy drop. 

\section{Conclusion}
In this paper, we propose the GnetDet model to optimize the object detection applications implemented on a low-power CNN accelerator chip. The proposed GetDet model takes practical factors into design considerations, including input format, model architecture, and output tensor. The experimental results show the joint performance of the proposed GnetDet model with minimum CPU load, fast inference speed, and excellent prediction accuracy is ideal for real-world applications, especially on the mobile and embedded devices.


{\small
\bibliographystyle{ieee_fullname}
\bibliography{egbib}
}

\end{document}